# FoldingNet Autoencoder model to create a geospatial grouping of CityGML building dataset

Deepank Verma, Olaf Mumm, Vanessa Miriam Carlow

Spatial Analytics and Cross disciplinary Experimentation Lab
at Institute for Sustainable Urbanism (SpACE Lab at ISU),
Technische Universität Braunschweig, Germany

{d.deepank, o.mumm, v.carlow}@tu-braunschweig.de

**Abstract:** Explainable numerical representations or latent information of otherwise complex datasets are more convenient to analyze and study. These representations assist in identifying clusters and outliers, assess similar data points, and explore and interpolate data. Dataset of three-dimensional (3D) building models possesses inherent complexity in various footprint shapes, distinct roof types, walls, height, and volume. Traditionally, grouping similar buildings or 3D shapes requires matching their known properties and shape metrics with each other. However, this requires obtaining a plethora of such properties to calculate similarity. This study, in contrast, utilizes an autoencoder to compute the shape information in a fixed-size vector form that can be compared and grouped with the help of distance metrics. The study uses "FoldingNet," a 3D autoencoder, to generate the latent representation of each building from the obtained LoD 2 CityGML dataset. The efficacy of the embeddings obtained from the autoencoder is further analyzed by dataset reconstruction, latent spread visualization, and hierarchical clustering methods. While the clusters give an overall perspective of the type of build forms, they do not include geospatial information in the clustering. A geospatial model is therefore created to iteratively find the geographical groupings of buildings using cosine similarity approaches in embedding vectors. The German federal states of Brandenburg and Berlin are taken as an example to test the methodology. The output provides a detailed overview of the build forms in the form of semantic topological clusters and geographical groupings. This approach is beneficial and scalable for complex analytics, e.g., in large urban simulations, urban morphological studies, energy analysis, or evaluations of building stock.

## 1. Introduction

Latent space corresponds to an abstract multi-dimensional feature space that encodes a representation of the original dataset (Foster, 2019). In other words, these are the compressed representations of the actual data points, such as an image, audio clip, etc., forming a vector space. It is effective for data analysis for two main reasons; first, it learns lower dimensions from a higher dimensional dataset while preserving most of the relevant information required to reconstruct the original data point. Secondly, it helps explore semantic sameness between data points, hence suited for studying clusters, outliers, and manifold learning. Latent space for complex datasets is most commonly obtained using autoencoders and Generative Adversarial Networks (GANs). Autoencoders are one of the unsupervised learning techniques that utilize neural networks to learn to efficiently compress and encode data. Various autoencoders use Convolutional and Recurrent Neural Networks (CNN and RNN) model architectures to encode image and audio datasets. As opposed to an autoencoder, GAN differs widely in architecture. Instead of encoders to compress the data, it utilizes latent space as a sampling space to generate image datasets. GAN-based latent space exploration (Härkönen *et al.*, 2020) has been quite prominent in manipulating image synthesis models that can create a variety of high-quality images (Zhu *et al.*, 2016).

Autoencoder-based encoded latent space applications include anomaly detection (An and Sungzoon, 2015), image and audio data denoising (Samal, Babu and Das, 2021), image inpainting (Yeh *et al.*, 2018), and information retrieval (Zhang, Liu and Jin, 2020). In recent years, autoencoders have been especially relevant in urban data processing. This is because many urban datasets, such as satellite and aerial imageries, datasets obtained from IoT devices, air pollution, and audio samples, are available as unlabeled datasets (Xiao *et al.*, 2022), suitable for unsupervised analysis. Similarly, the 3D model of buildings is one such urban dataset that has been widely studied lately as it provides insights into the types and locations of human dwellings and workplaces more clearly than its 2D representation as footprints.

Data related to buildings, especially in cities, is often well maintained in cadaster for planning and administration, census and revenue purposes. Even where such information is unavailable, Deep Learning (DL) methods focusing on building detection and identifying footprints (Vakalopoulou *et al.*, 2015) have made this problem trivial. These methods have been applied to detect footprints from dense areas to remote places (Microsoft, 2023). Apart from 2D footprint mapping, governments worldwide have started preparing cadaster of 3D buildings using OGC CityGML (Kolbe, 2009) format and making such data accessible within the framework of open data policies. 3D built form constitutes geometric information regarding building footprints, roof shapes, walls, height, windows, and doors; therefore, it is interesting from a research point of view in architecture and urban design, demographics, and disaster prevention. Smart cities and their management consistently demand updated city visualization, preferably in 3D, for which researchers have used multiple approaches to utilize aerial Lidar datasets (Park and Guldmann, 2019; Nys, Poux and Billen, 2020) to generate 3D city models.

Recent developments in DL methods have also renewed the interest in 3D building modeling approaches such as façade reconstruction (Bacharidis, Sarri and Ragia, 2020), 3D building reconstruction from street view images (Pang and Biljecki, 2022) and aerial images (Alidoost, Arefi and Tombari, 2019), and building cognition and shape coding (Yan *et al.*, 2021). Studies (Tutzauer *et al.*, 2016) have also investigated perception-aware visual communication of 3D buildings, which defines a human cognitive experience. Recently, conscious efforts have been made to understand the inherent characteristics of the 3D shapes of buildings. Labetski *et al.*, (2022) created a software package to compute 2D and 3D building metrics, which can be used to generate building clusters and understand urban morphology. Similarly, Yan *et al.*, (2021) used a Graph convolutional encoder to learn building shape representation to aid cognition of buildings in maps.

The development of buildings over the years, the changes in building designs and characteristics due to socio-economic changes, and their locations in the urban fabric potentially provide a great evolution story of settlements. The buildings and their inherent characteristics, such as structure, architecture, type, and function, have long been studied to assess the aesthetical and structural design and floor plan layouts. However, such studies are not focused on detailing and generalizing individual phenomena at larger scales involving multiple cities and regions due to availability issues of relevant datasets. Building grouping has been attempted to generate insights and extract geographic patterns showing morphological variations (Yan *et al.*, 2022). These groupings are especially relevant for conducting simulations and large-scale aggregations. For instance, clustering buildings based on their design/archetype has excellent potential for generating a quick understanding of energy consumption (Li *et al.*, 2018). With the availability of large 3D datasets, readily calculated features such as footprint area, height, and volume offer quick information on building types. Further, recently developed 3D building metrics (Labetski *et al.*, 2022) provide many 3D shape metrics that can be utilized to compare and group any set of buildings.

However, rather than utilizing metric methods, this study considers building grouping an unsupervised classification problem and uses latent-based data analysis to uncover the inherent relationships between the buildings. Two main reasons to choose the unsupervised method over metric-based methods are (a)

Semantic representation and generalization: latent embeddings capture semantic information about the 3D shapes, which goes beyond basic geometric properties, which further implies that embeddings can encode information about the shape's structure, orientation, and more abstract features. Additionally, they can capture similarities and differences between shapes that might not be evident from simple metrics. This enables the model to generalize across a broader range of shapes and recognize shared characteristics, making it more versatile in handling variations in shape. It is also well-suited for more complex shapes, such as LoD greater than 2.0, and/or 3D shapes other than buildings, and (b) DL-based classification tasks: the embedding model trained on a large dataset of 3D buildings can be fine-tuned for specific tasks, such as classification or regression. These models can use transfer learning to deploy already learned characteristics to new 3D data sets.

However, learning appropriate representation from 3D shapes has been a fundamental problem in shape-based recognition, clustering, and classification (Osada *et al.*, 2002). Due to the arbitrary types in 3D datasets, non-fixed orientations, and sizes, the methods used in traditional 2D and text-based datasets are inapplicable. Historically, multiple strategies have been implemented to determine ways to measure 3-D shape similarity, such as probability distribution, 2D image representation, Contour and Fourier descriptors, and comparison of high-level descriptors of common 3D shapes (Basri *et al.*, 1998; Alt and Guibas, 2000; Osada *et al.*, 2002). Recent developments in 3D-specific DL models have addressed the issue with varied outcomes.

This study utilizes the CityGML-based LoD 2 building dataset and a 3D point-based autoencoder to generate latent vectors, which are further used to investigate (a) latent space representation and exploration of data clustering in latent space, (b) assess the quality of reconstructed buildings, interpreting cluster semantics by hierarchical clustering, and (c) discuss practical applications such as geospatial grouping to understand the location of similar buildings and building groups. These groupings provide an overview of settlements' structure and potential relationships in the geospatial domain.

## 2. Methodology

### *2.1. Data Sources*

The northeastern German federal states of Berlin and Brandenburg are selected in the study (Figure 1). The size of Brandenburg is approximately 29,650 sq. km, divided into 413 municipalities in 18 counties and county-free cities. Similarly, Berlin is approximately 890 sq. km divided into 12 districts and 103 sub-districts. The geoportal database of each state, LGB (LGB, 2023), and FISBroker (Geoportal Berlin, 2023b) provides access to geospatial datasets, including historical imagery, high-resolution orthophotos, and LoD 2.0 GML datasets arranged in a tiled manner. For this study, approximately 16,500 tiles for Brandenburg (GeoBasis-DE/LGB, 2023) and 900 tiles for Berlin (Geoportal Berlin, 2023a) are downloaded, consisting of 2.32 million LoD 2 buildings and 0.65 million for Brandenburg and Berlin, respectively. LoD 2 data for Brandenburg was created on 31.12.2012 and was last updated on 17.05.2022, whereas data for Berlin was created on 01.04.2022 and recently updated on 10.05.2023. As the capital city, the building number density is significantly higher in Berlin at 756 buildings/sq.km, followed by the county-free cities Cottbus, Potsdam, Brandenburg an der Havel, and Frankfurt (Oder) in Brandenburg state with 304, 262, 201, 174 respectively. In contrast, border counties such as Uckermark, Prignitz, and Ostprignitz–Ruppin have the lowest density of 36, 43, and 46, respectively.

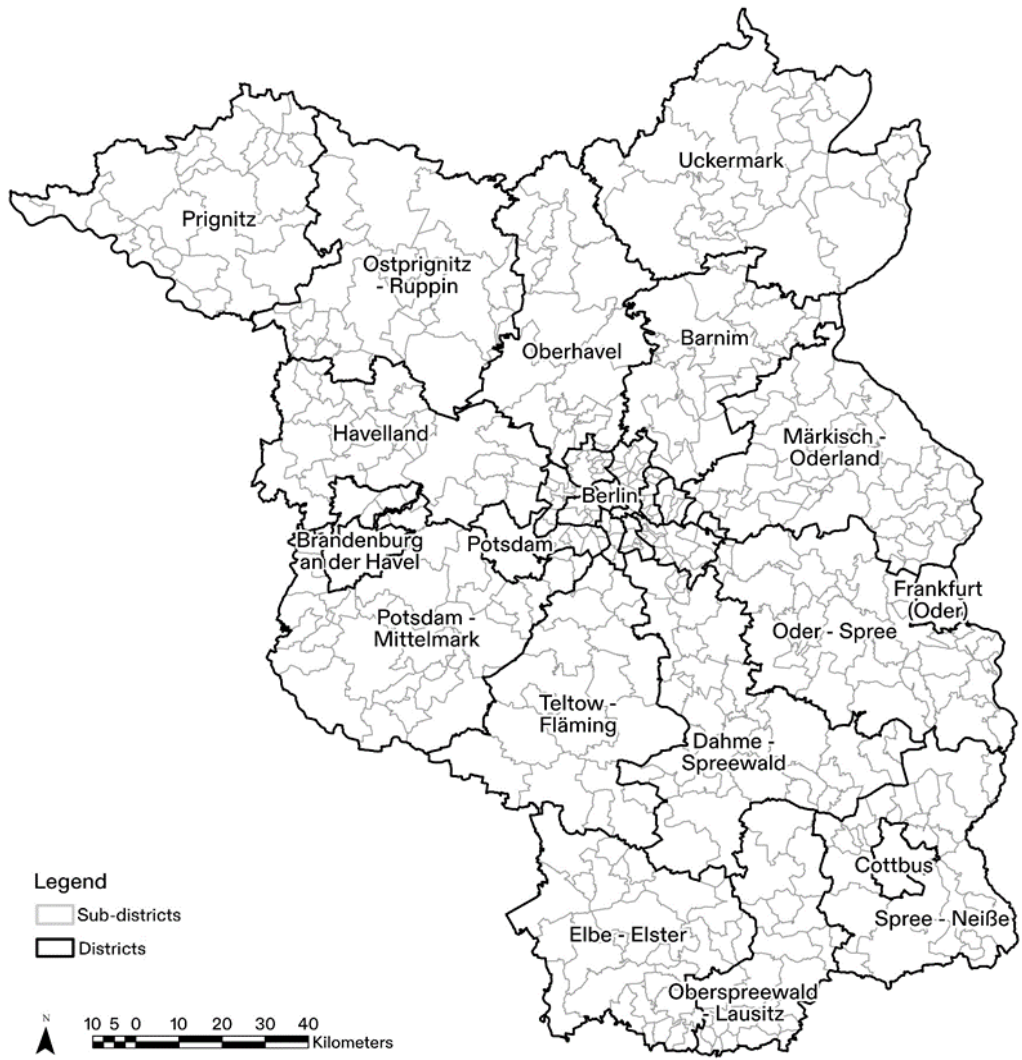

*Figure 1 Map showing the study area of Brandenburg and Berlin*

For an easier interpretation of administrative boundaries throughout, we refer counties and county-free counties as districts and the municipalities as sub-districts in Brandenburg, similar to the convention followed for Berlin.

## 2.2. Overall Methodology

The overall methodology (Figure 2) includes the (a) preprocessing of the CityGML dataset, including removing structures other than buildings. As this process relies on buildings being watertight mesh, buildings are also removed that are not watertight. The remaining meshes are converted to an *OBJ* format for better visualization and downstream processing. (b) The *OBJ* model is converted to a point cloud using sampling methods and fed into the FoldingNet autoencoder for training. (c) the generated embeddings are then used for visualization, to create and subsequently analyze the quality of reconstructions of the buildings, and to produce clusters using hierarchical clustering methods. Finally, (d) embeddings are also used to develop a geospatial tool to understand building groupings in the selected area.

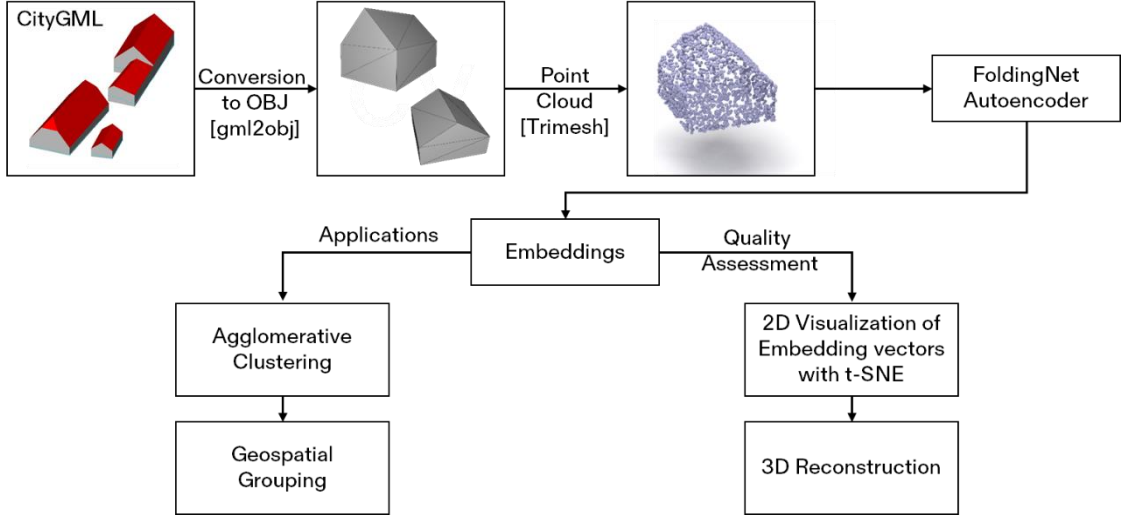

*Figure 2 Flowchart showing overall methodology*

## 3. Latent feature extraction

Autoencoders can process large text, audio, video, and image datasets to compute latent spaces. However, such methods are not directly applicable to 3D datasets. For example, in contrast to RGB matrix as an image with fixed and consistent pixel indexing, 3D shapes have a non-fixed orientation and have various extents. Therefore, standard DL encoding models cannot learn these variations effectively and have not been prominent in utilizing 3D datasets. To remedy this, researchers have undertaken methods to simplify shapes to voxel grids (Wu *et al.*, 2015) and point clouds (Qi *et al.*, 2016) for classification and representation learning.

Recently, these methods have become more prominent in Simultaneous Location and Mapping (SLAM) setups. 3D point cloud datasets have been collected in large amounts as Lidar datasets from satellite/aerial platforms and driverless vehicles that employ scanners as part of their control and obstacle avoidance system. Although these systems utilize sensor fusion for comprehensive perception and increased accuracy and reliability, Lidar-based point clouds and their classification are essential in detecting pedestrians, vehicles, and street curbs. Models such as EdgeConv (Wang *et al.*, 2019), PointNet++ (Qi *et al.*, 2017), and KPConv (Thomas *et al.*, 2019) have been prominent in point cloud classification. Labeled datasets such as ModelNet (Wu *et al.*, 2015), ShapeNet (Chang *et al.*, 2015), and SemanticKITTI (Behley *et al.*, 2019) have been used by researchers to test new classification models. Since point clouds are notoriously difficult to label and require much expertise and human hours (Xiao *et al.*, 2022), the openly available unlabeled point cloud datasets far outweigh the labeled ones such as these.

Unsupervised learning shows prominence in point cloud data exploration and understanding. These methods can be utilized for feature learning and representation using autoencoders. Domain adaptation and transfer learning can be used for classification in the domain where the labeled datasets are scarce.

### 3.1. FoldingNet

Classifying irregular structures of sparse 3D points has been difficult with the regular DL frameworks since CNN requires its neighboring samples at fixed distances to facilitate convolutional operations, while point datasets do not have such a fixed structure. Multiple strategies such as Voxelization (3DCNN), image-based descriptors (Chen *et al.*, 2003), and multi-view 2D images (Su *et al.*, 2015) have been utilized to alleviate the problem; however, they provided less feasible results. PointNet (Qi *et al.*, 2016) architecture overcomes this shortcoming with the symmetry function, which makes the point cloud invariant to the input order and thereby removes the compulsion to input the cloud in a specific order in the model. FoldingNet (Yang *et al.*, 2018) is an end-to-end deep autoencoder that follows PointNet architecture designed to address unsupervised learning challenges in point clouds (Figure 3).

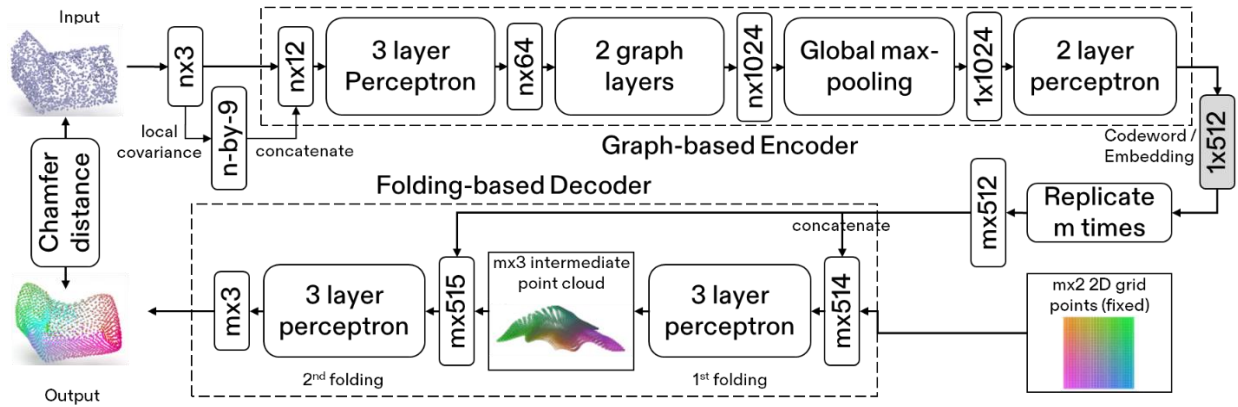

*Figure 3 FoldingNet architecture (Yang et al., 2018)*

From a general point of view, the FoldingNet is a regular Autoencoder with Encoder and Decoder halves separated by the bottleneck layer. It primarily involves a 2D points plane that learns a folding operation, like cutting, squeezing, and stretching, through training and subsequently reproducing the 3D point samples as part of Decoder architecture. The output from the bottleneck layer is the latent vector of size 1x512, called codeword in the FoldingNet model. The input to the encoder is the nx3 matrix, where n represents the number of points in the point cloud. The encoder is based on multi-layered perceptrons and graph-based max-pooling layers. FoldingNet uses a 2D grid plane consisting of points, and the codeword obtained from the encoder and uses two consecutive 3-layer perceptrons to fold the 2D grid into the input point cloud shape. The reconstruction error is computed by chamfer distance, which is back propagated in the model. In the initially proposed model, FoldingNet is further used for semi-supervised learning and transfer learning with the help of obtained codewords. This study utilizes the FoldingNet model to obtain codewords (embeddings) from the 3D shapes of the buildings for reconstruction and latent space visualization to assess the quality of the trained model.

### 3.2. Processing CityGML dataset

CityGML is an XML-based standard OGC format for storing 3D city models into meaningful semantic parts such as roofs, doors, windows, and interiors. Apart from buildings, CityGML is also capable of storing details such as tunnels and bodies of water. It provides five standard Levels of Details (LoDs) from 0 to 4, where 0 is a 2D building footprint, 1 is a block model obtained by using height for extrusion, 2 includes enriched surfaces with roof design, 3 includes door and window details, and others fenestration, and 4 consists of the interior details in the building (Kolbe, 2009). Although CityGML is a feature-rich model standard with the potential of many scientific applications, it is not as widely adopted as other formats such as OBJ and FBX. Due to this, not many standard 3D programs support CityGML (Biljecki and Ohori, 2015).

As GML files are essentially XML files, they can be parsed to extract other information attached to them. The GML dataset downloaded for this study comprises roof type, building function, and height information. Each structure has one of the 15 roof types and 302 classes of building function, defining the current use of a particular building. The height of the building is given from the ground to the highest point in the structure. The mean height of buildings in Berlin is 6.7 m with a standard deviation (s.d.) of 5.6, while the mean is 4.5 m and s.d. of 3.6 in Brandenburg. We also analyzed the roof type and function information from all the buildings in the study area. Berlin has around 42 % of the buildings with flat roofs and 21 % as pitched roofs, whereas Brandenburg has 33% each flat roofs and pitched roofs; other roof types include pent-roof (8.6%) and hip-roofs (2.1%). Further, Berlin has around 42% of buildings as residential houses, followed by canopy, shed, and garage combined at 31.18 %, while commercial buildings at 6.71%. Brandenburg comprises 31.7% of buildings as residential houses, around 14% as garages, 9.7% as agricultural and forestry buildings, and 9% as recreational buildings.

Each building structure is individually extracted from the downloaded GML tiles, which are then converted to *OBJ* format with the help of *Citygml2obj* (Biljecki and Ohori, 2015) for better visualization and data interpretability in the standard 3D softwares. However, the format conversion introduces errors which are due to the presence of non-watertight meshes or holes in the surfaces. The frequency of such errors is significant and causes around ~30 percent of the GML shapes to be rendered error-prone on conversion. The study removes these shapes from further consideration, bringing the building dataset's effective size to ~2 million. Further, to ease computation, we randomly selected 25% of the remaining buildings for model training instead of all the obtained buildings. These 25% (0.5 million) buildings are divided into 3:1 to create training and testing sets.

The *OBJ* models cannot be directly used as inputs to the FoldingNet model; hence, each obtained building is converted to a point cloud of 2048 points with the help of the Trimesh library (Dawson-Haggerty et al., 2019) in Python using the sampling function. The process takes random points from the faces' surface in the 3D model. Similar studies (Qi *et al.*, 2016; Yang *et al.*, 2018) utilizing the 3D Autoencoder method use ShapeNet (Chang *et al.*, 2015) and ModelNet (Wang *et al.*, 2019) datasets and normalize the point cloud in unit sphere dimensions after sampling to facilitate efficient computations. In contrast to these curated datasets, built structures have significant variations in shapes and sizes. Therefore, converting all the built point clouds to unit sphere dimensions as in the original FoldingNet model would not preserve the height and volume of the shapes, which is relevant in this study. However, given the range of dimensions in buildings, considering the whole set of build-dataset without proper scaling will produce convergence difficulties in model training. To solve this issue, the distance between the centroid and the farthest point on the cloud of each building is calculated, after which the buildings falling within the 1-99 percentile distance range are kept and then normalized. The outlier buildings in this process are removed. The obtained building points dataset are then fed to the FoldingNet.

### 3.3. Model Training

A python-based version (Tao, 2020) of FoldingNet architecture is utilized in this study without significant hyperparameter modifications, except for the number of epochs, learning rates, and batch sizes. A total of six instances of the model are trained by varying the size of the bottleneck layer from the original 512 to 256, 128, 64, 32, and 16. Changing the bottleneck size is experimented with to assess the degree of generalization even with fewer embeddings. Each model is trained for 800 epochs with an ADAM optimizer, with an initial learning rate of 0.0001 and a batch size of 16 (Figure 4). The optimum batch size and learning rate were investigated before selecting the current ones.

The loss curve for each model training exhibits a rapid fall followed by a slow and gradual decrease throughout training without apparent plateau or stagnation. This effect might be due to the data complexity of the buildings with intricate patterns and the selection of smaller batch sizes, which induce more noise, leading to slower convergence. It can be well established that decreasing the codewords results in higher reconstruction loss, which is well reflected in Figure 4. However, the differences in losses are gradual, except for the model with 16-D embeddings (M16). Given extended training durations, the M256 and M128 models can potentially generate analogous reconstruction losses to the M512 model.

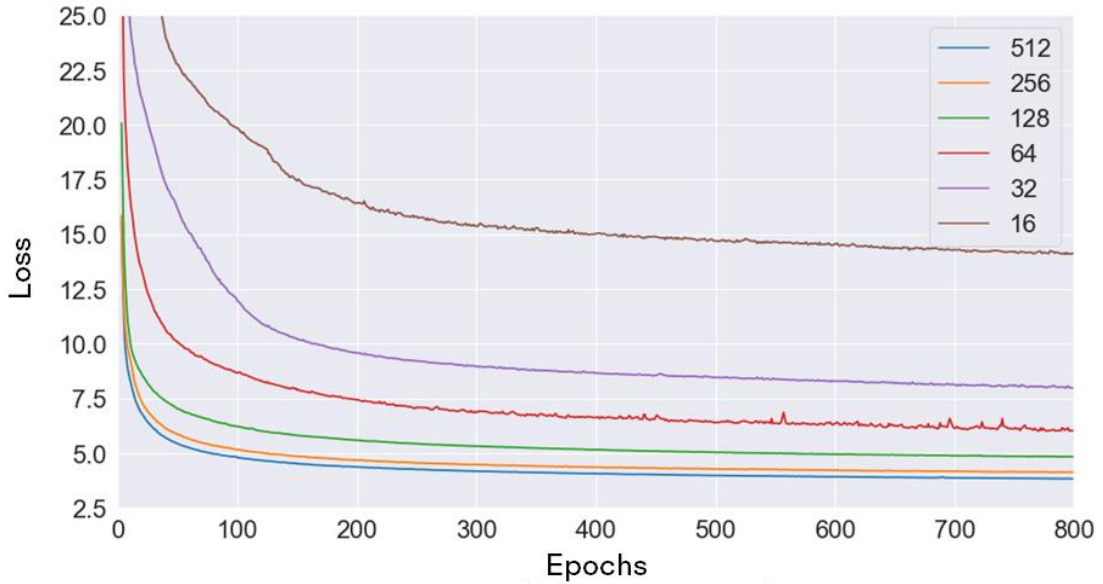

*Figure 4 Reconstruction loss for the models with varying embedding sizes.*

### 3.4. Results

We randomly chose 100,000 buildings from Berlin to generate embeddings and subsequent reconstructions using six trained models. The model reconstruction results show the capability of trained embeddings from six different bottleneck layers. The decoder can generate an original point cloud (Figure 5) for almost all types of buildings. As seen in model training, the reconstruction quality depends widely on the dimensions of the bottleneck layer. The M16 model struggles to reconstruct the input data with similar accuracy to large bottleneck layers. While the size and overall cloud faintly represent the buildings in smaller dimensions, the reconstructed point clouds are distorted with no precise edges formed, giving no clear indication of the type of building being reconstructed. It still preserves the overall shape of the building in the reconstruction but loses much finer details. The quality of the reconstructed cloud increases significantly with larger embeddings. The edges can be seen taking shape at M32, which gets more refined in M64. M128 and M256 show improvement in creating significant edges and proper footprint and walls; however, the roof structures are still less formed. M512 shows an almost 1:1 replica of the input point cloud. As consistent with the reconstruction loss (Figure 4), the model's capability to reconstruct the data jumps significantly from M16 to M32, while only slightly incremental quality improvements from M128 to M512.

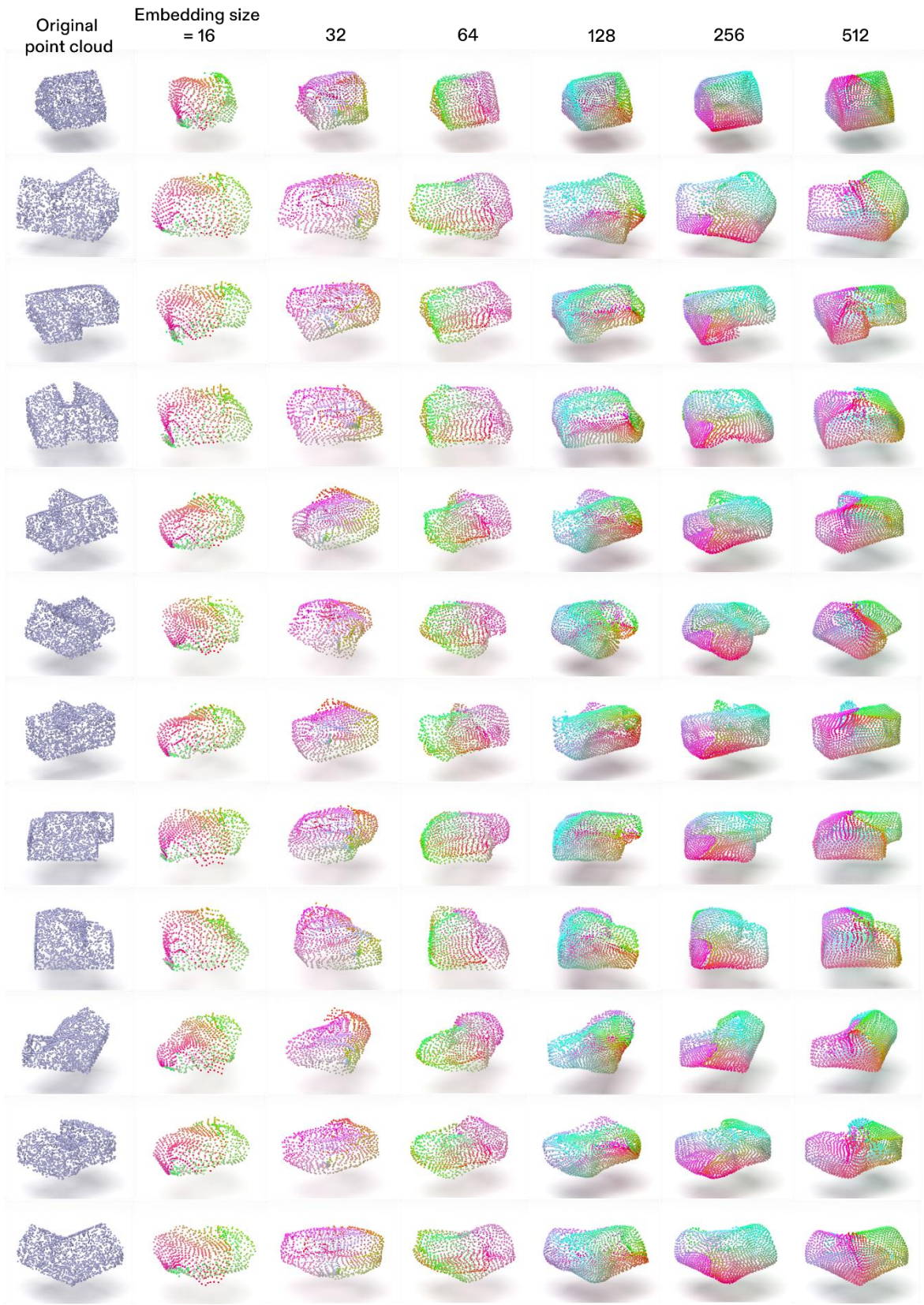

*Figure 5 Reconstruction outputs from the models with varying embedding sizes*

The obtained embeddings for each building structure from the M512 model are then utilized to generate 2D representations using t-SNE (t-distributed Stochastic Neighbor Embedding) (Figure 6). A t-SNE model primarily visualizes high-dimensional data in a lower-dimensional space while preserving local structures and similarities between data points. It is a non-linear visualization method in which the distance between points on the plot does not correspond to the original high-dimensional space. We chose the perplexity value of 80, the learning rate in 'auto' mode with 1000 iterations to produce the results. Figure 6 shows an interactive t-SNE plot created using the *Bokeh* Python library; the popups show the nearest neighbors and their *gml* ids near the mouseclick. The role of the autoencoder in capturing meaningful representations of the data is visible from the plot. The data points appear densely packed and spread throughout the visualization space. It is a common phenomenon for complex and high-dimensional datasets. The specific formation of small tight clusters is also due to the t-SNE, which focuses on preserving local structures rather than global ones.

We examined the small clusters and scattered data points spread across the space. To analyze the distribution, we selected 12 points from the spread. This enables us to visualize the local neighbors and ascertain whether they exhibit gradual variations along the latent space or form local clusters. Cluster 1 is mainly dominated by extremely tall structures, while 2 shows horizontal plane-like structures different from regular cube-like structures representing buildings. Upon investigation, it shows that these structures are outliers in the dataset as they have building functions erroneously listed as warehouse storage, shed, and residential building. The clusters are selected to show an increasing order of complexity in the number of faces, roof types, and sizes. However, the complexity of shapes does not increase linearly; they are distributed in relatively small clusters and/or spread. Small and tighter clusters such as 11 and 12 indicate built structures with exactly the same type.

Although the t-SNE plot does not show distinct clusters, the ability of the autoencoder to compress and reconstruct the data effectively suggests it learned valuable features and patterns. Since t-SNE is best for visualizing the data spread, it is not viable to use it for clustering due to its stochastic nature and preservation of local similarities to global ones; we utilized hierarchical clustering to understand building grouping and comment on the formed cluster type.

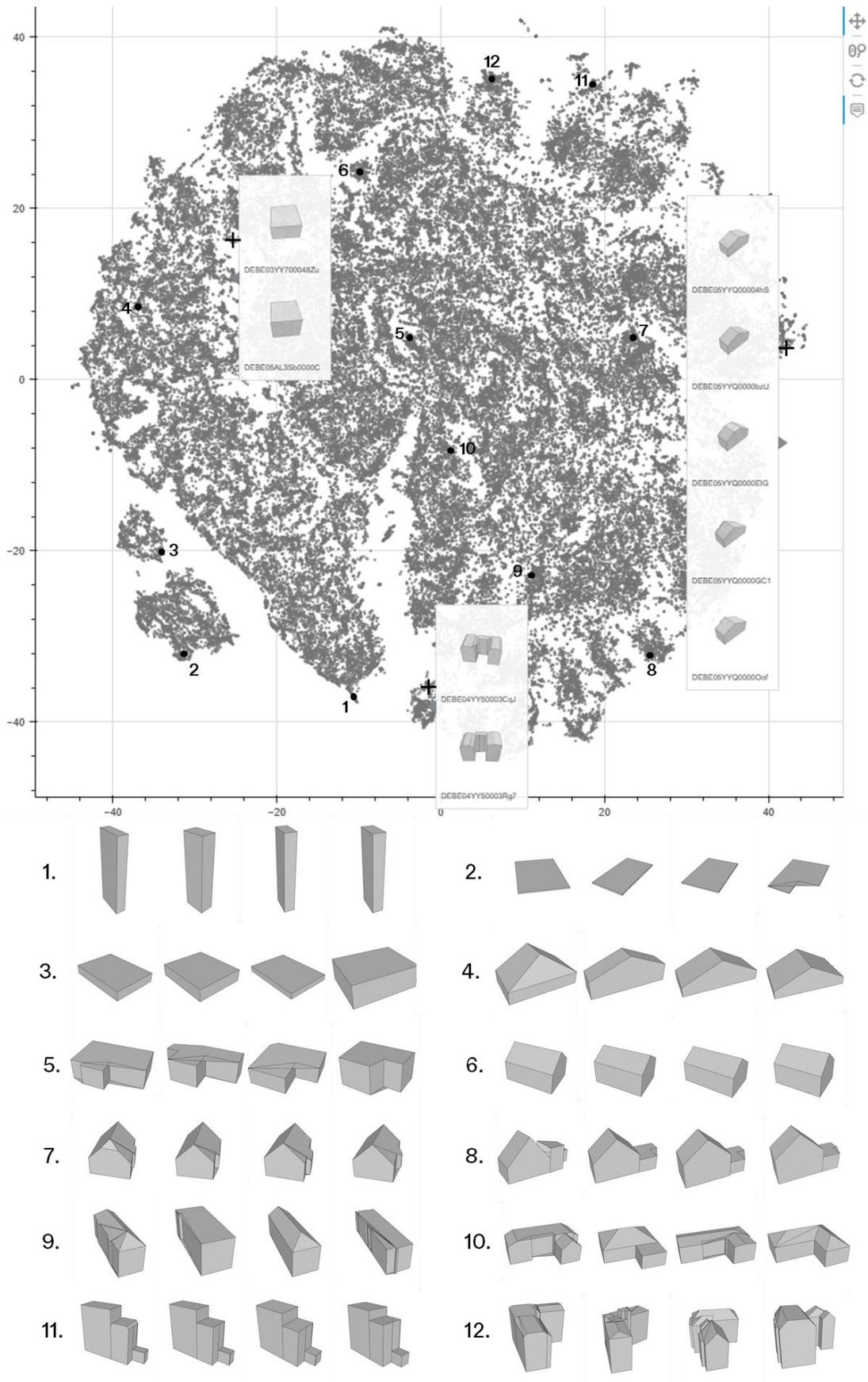

*Figure 6 t-SNE scatterplot of embeddings obtained from the M512 model and the nearest neighbors of the selected locations*

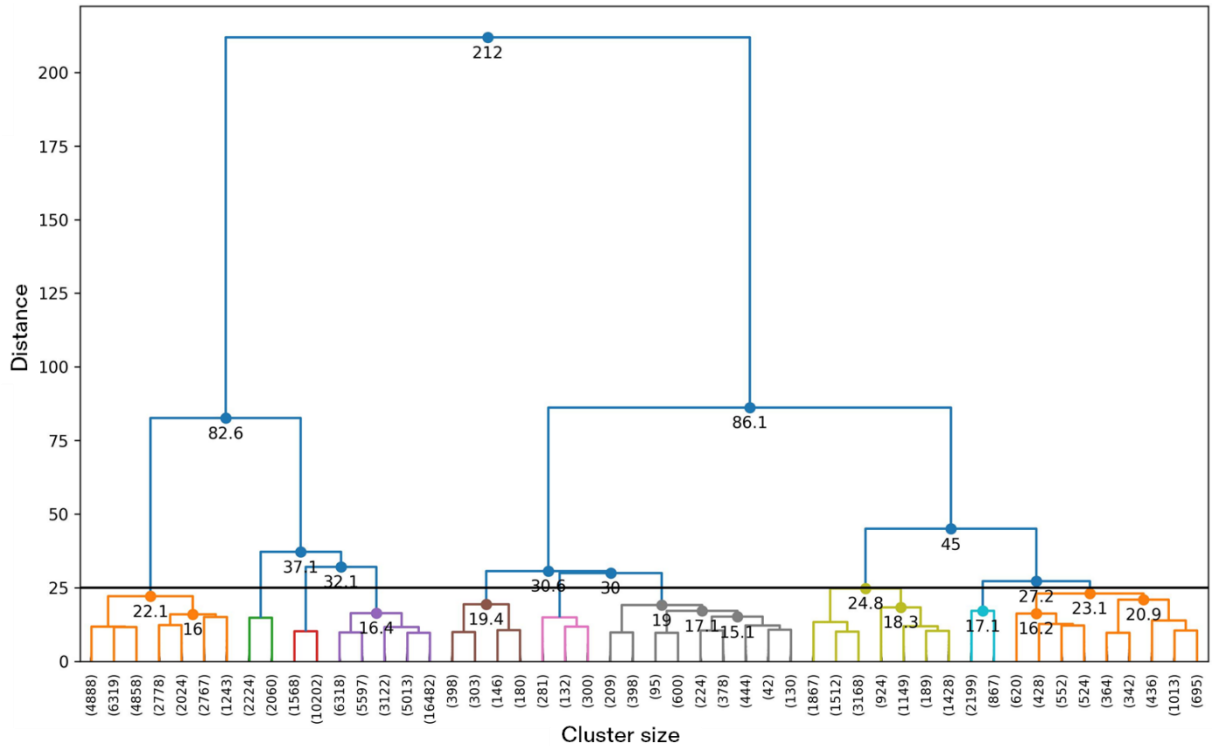

*Figure 7 Results of agglomerative clustering on embeddings obtained from the M512 model when choosing cutoff at d=25 to form 10 clusters*

Hierarchical clustering can be executed with agglomerative (bottom-up) and divisive (top-down) approaches. In this study, we used the agglomerative method, where each data point starts as its own cluster and iteratively joins the closest clusters to create a single cluster. We used the *scipy* library to create hierarchical clusters using *ward* linkage to determine the distance between clusters. The embedding values with 512-D are reduced to 15 features using the PCA approach before calculating clusters. Figure 7 shows the dendrogram, which represents relationships and similarities between data points. The vertical axis shows the Euclidean distance, while the horizontal represents each data point/cluster. It further provides information on the number of data points within each cluster and their distance from the main branches. Since the embedding spread does not show clear clusters, the decision was taken to use different cutoff distances to visually investigate if sufficient differences between the structures in clusters are visible. The dendrogram shows that the embeddings form 4 clear clusters at *d=82.7*, which branch out to 10 clusters at *d=25*. However, upon investigation, the 4 clusters show muted differences in their structures simply because each cluster covers a large chunk of samples, which does not give a clear difference in built structures within them. We chose a dendrogram cutoff distance of 25, which produced 10 clusters. We randomly sampled 12 buildings for each cluster obtained with the selected cutoff distance.

The differences between the 10 clusters (Figure 8) are not especially stark; however, it gives an overview of how the clustering algorithm has grouped the buildings based on their structure, volume, roof types, and number of faces. Clusters 1 and 3 mainly show a mix of high, mid, and low-rise structures with linear shapes, while cluster 4 shows large horseshoe/U-shaped structures slightly similar to cluster 2, with mixed structure designs including U-shapes. Clusters 5 and 6 show simple structures with fewer faces and mostly flat roofs, while cluster 10 also shows simple shapes but with non-flat roofs. Similarly, cluster 8 shows simpler cuboid (rectangular/boxy) forms, significantly differing from cluster 7, which shows more complex built structures. Cluster 9 shows tall and slender buildings, representing skyscrapers.

While our analysis delved into the clusters obtained, it is essential to note that the clustering was performed on a subset (100,000 buildings) of a larger dataset (~2 million). Incorporating the entire larger dataset would yield different clusters and their distributions due to increased scope and variety in the dataset. Despite the expected dissimilarity in clusters when considering a larger dataset, it is worth emphasizing that the analysis conducted on this smaller dataset offers a valuable overview of the capability exhibited by the autoencoder. This primary exploration has provided insights into underlying structures and patterns within data.

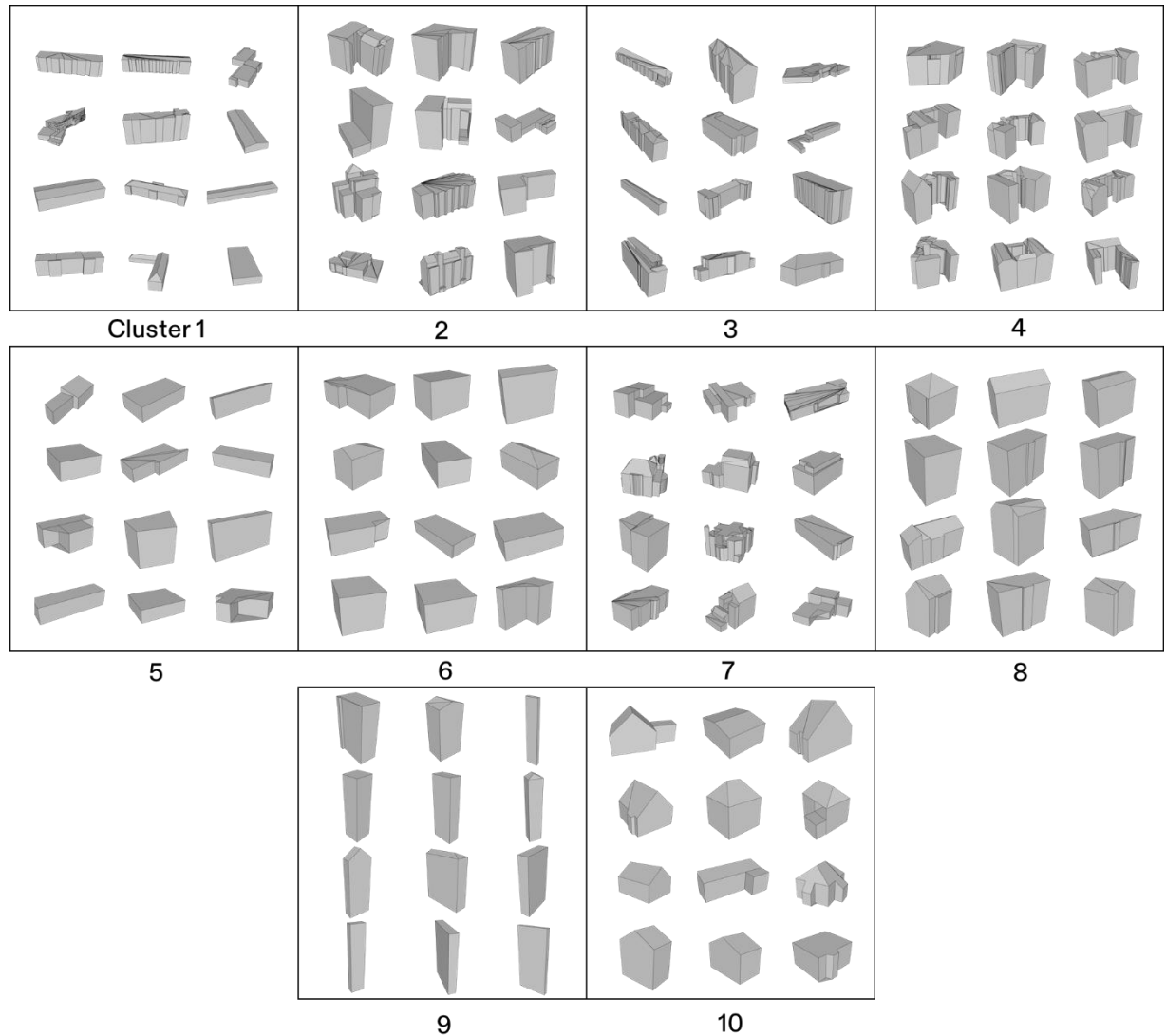

*Figure 8 Random samples of buildings within each of the 10 clusters obtained from agglomerative clustering*

## 4. Geospatial grouping

While the embeddings obtained from encoding the 3D buildings and their visualization can provide information regarding the tight clusters and patterns emerging from the latent space spread, it does not give enough information on the geospatial grouping of the buildings. Since the locational coordinates of the buildings are already known, translating the 2D spread into the geospatial map is just a trivial task. Creating a geospatial grouping through this method is, therefore, more straightforward if the clusters are already formed and subsequently labeled as such. However, since we did not delineate clusters and groups in latent spread manually or otherwise with other clustering algorithms, translating the space into geospatial locations will not be geographically clustered. In short, the clusters in latent space would not necessarily form clusters in the geospatial domain, where the points from the same groups might be scattered around the study area.

We developed a model to integrate latent distances with the geospatial to group similar buildings. For this exercise, the embeddings from the 512D model for each of the ~2M buildings are generated. Since the selected study area is vast, the context of the groupings is restricted to the smaller administrative boundaries, i.e., sub-districts in Brandenburg and Berlin. The geospatial modeling process (Figure 9) includes (a) identifying the median center of each administrative unit (total = 516 units). For this process, the center would act as an anchor point and de facto center for that unit. (b) The nearness (in meters) of each building to this center is determined with the help of the "*Near*" geostatistical tool in ArcGIS. Further, (c) the iterative process assigns the nearest building = group 1 and finds the nearest neighbors of the selected building based on Cosine distance (n). (d) all the identified n neighbors are assigned group 1 and are locked in further iterations. (e) Consecutively, the process selects the second nearest building and repeats steps c and d until every building is assigned a group number.

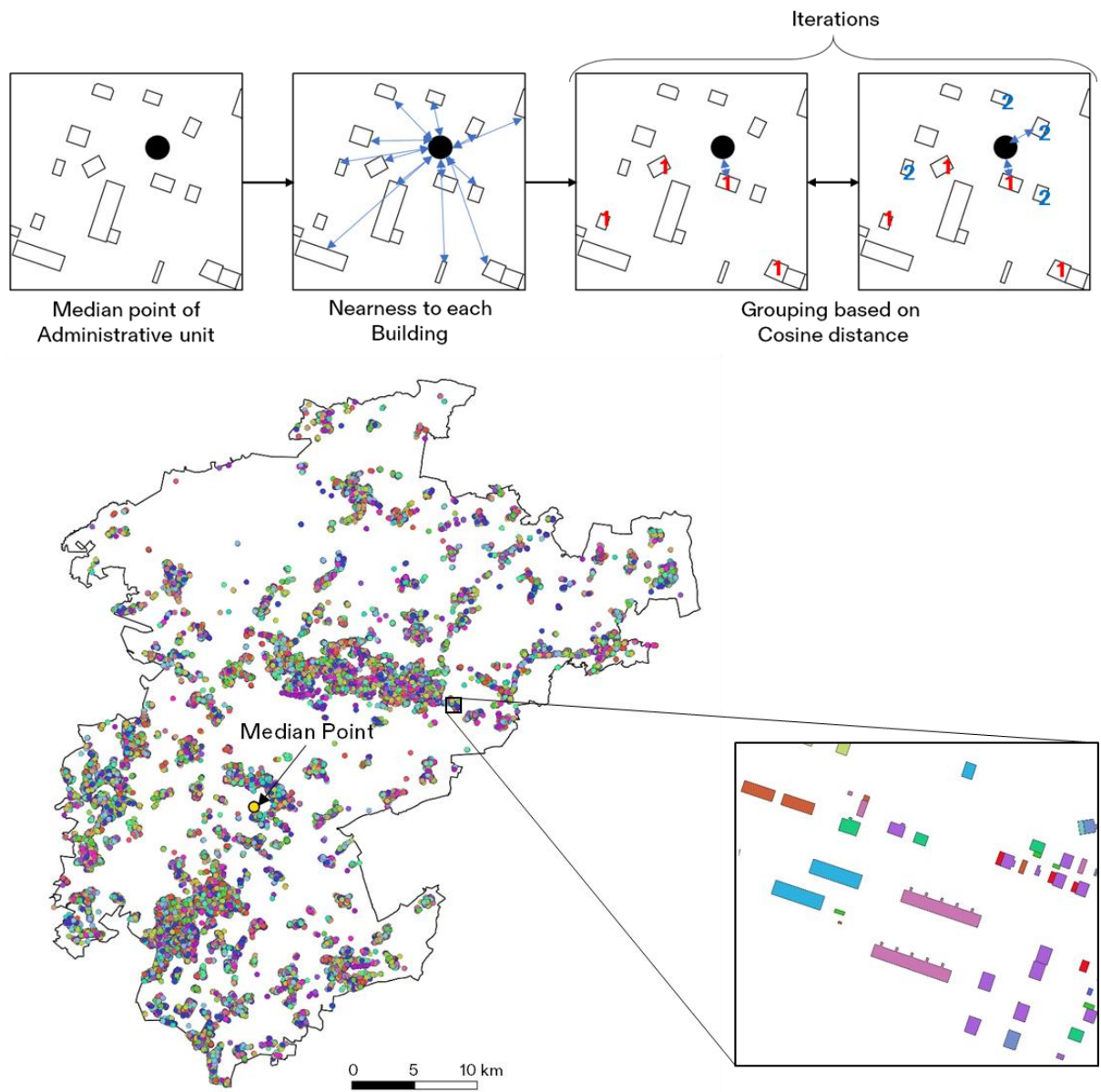

*Figure 9 Geospatial grouping strategy based on cosine distance (n)*

The number of groups formed depends on the value of 'n,' cosine distance. The value 'n' is selected manually while iterating and evaluating the results over all the potential values, such as 0.01, 0.02, 0.03, 0.04, and 0.05. The higher the value, the bigger clusters are created, which in turn include not-so-similar buildings. In contrast, the smaller value makes many overly small groups with more similar buildings. We chose the value of n = 0.03 in this study, where we felt it was a good compromise to select the optimum number of points in the clusters while ensuring the similarity of buildings.

The output of this process can be seen in Figure 9, in which points in the map represent each building where the color suggests the group. Due to the vast diversity of building shapes and sizes, no clear pattern can be seen on the main map; however, similar buildings being classified as a group can be seen by zooming. Similar structures are seen to be grouped together; however, to properly comment on the result, actual shapes must be visualized rather than just the footprint. This method also allows one to choose settlement boundaries with some underlying context. For instance, the model can benefit by utilizing settlement boundaries, calculated by considering the indicators of form, function, and spatial linkages such as area, compactness, building density, functional richness, population density, etc. (Carlow *et al.*, 2022).

The process is repeated for each sub-district (Figure 10), wherein each sub-district consists of every building assigned a group obtained through the above geospatial model involving the similarity values and number of buildings. Given the same number of buildings in two sub-districts, more groups in one can mean more variety in building types and vice versa. However, since the number of buildings varies widely in these sub-districts, calculating variety through this metric alone is impossible. Taking the ratio ($k$): total number of buildings / total number of formed clusters provides a better metric for estimating the variety of building types. The lower value of $k$ implies more diversity and vice versa.

Figure 10 shows the ratio ($k$) representation in all the sub-districts. The diversity of built forms is evident in Berlin as the city's sub-districts offer smaller $k$ than the adjoining sub-districts of Brandenburg, which form a distinct urban fringe around Berlin state borders, implying inherent homogeneity in building typology in these sub-districts. Apart from this difference, no visible pattern shows diversity and monotonicity among other sub-districts of Berlin and Brandenburg. However, it can be seen that the sub-districts with smaller areas, including sub-districts in Brandenburg state, generally have smaller $k$ than the larger ones. This is explainable, at least rudimentarily, since administrative regions are defined by population, which in turn refers to a larger density of the number of buildings. The denser urban area further implies diverse buildings, which might be true for older urban cores, e.g., in Berlin; hence, the ratio is lower for smaller sub-districts. However, many Brandenburg districts are smaller and have fewer villages with a lower $k$. This can be further explained as these small villages include various build types such as warehouses, silos, sheds, and residential houses, which adds to the diversity. The opposite is also true for larger sub-districts with higher $k$ as these sub-districts have the same build types but with larger numbers, which ultimately accounts for a larger value of $k$ than the smaller regions.

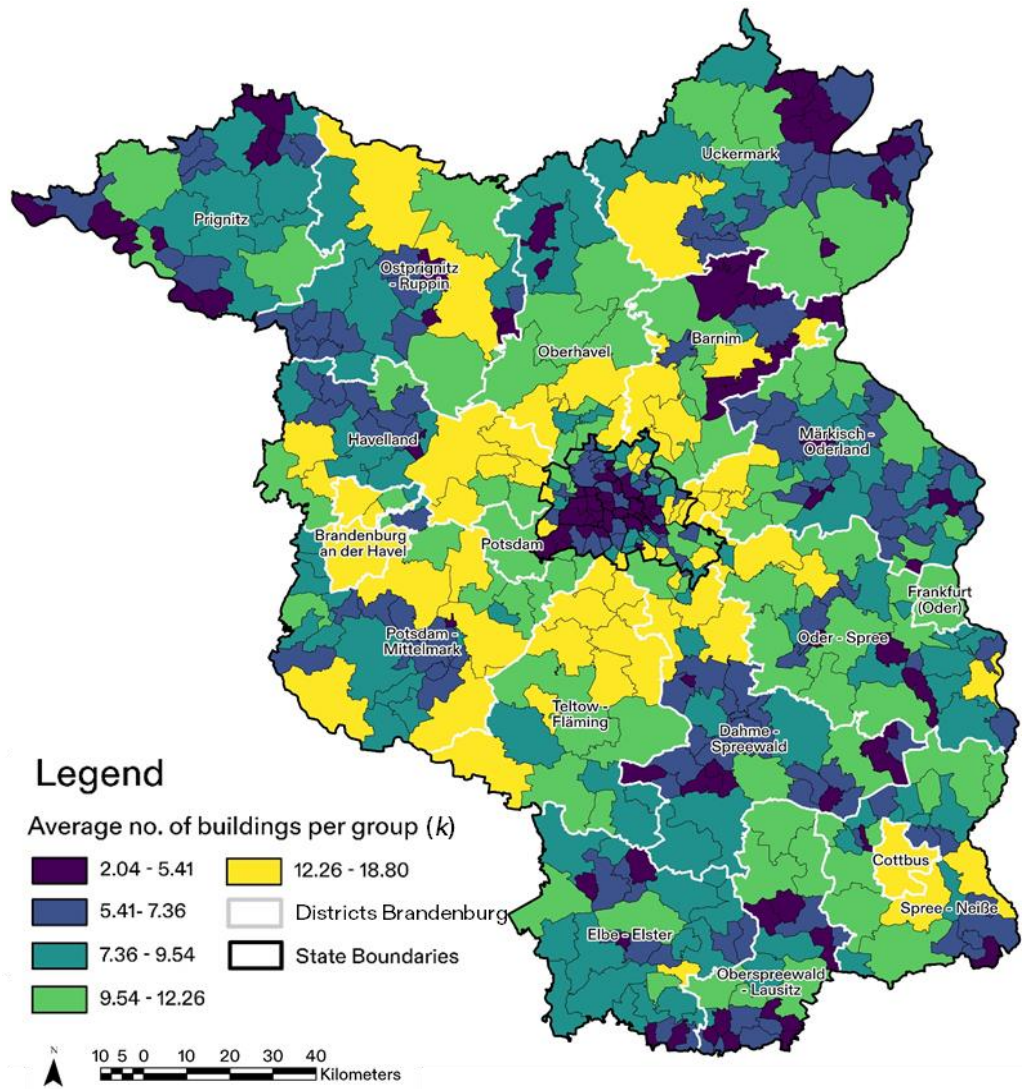

*Figure 10 Map showing the outputs of the grouping strategy conducted for all the sub-districts in the study area*

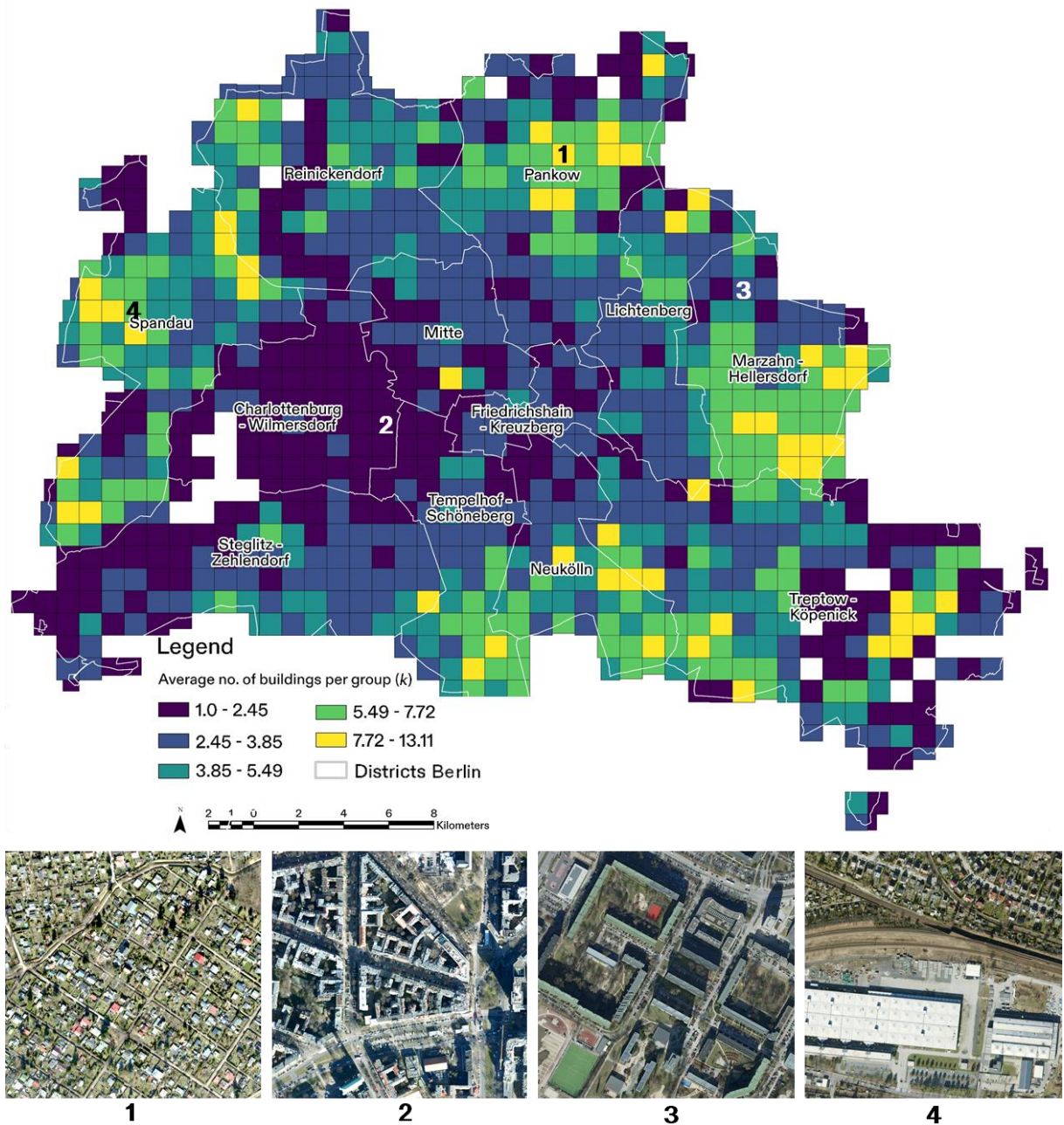

*Figure 11 Outputs from the grouping strategy conducted on a tiled map of Berlin (above). Zoomed in images for selected tiles (below)*

Although the building groupings according to the districts provide a visual comparison of the city's overall structure, it does not detail enough to comment on local distributions of the buildings. To understand that, we created window tiles of 1km x 1km to cover Berlin (Figure 11), which now act as a boundary instead of sub-districts. The exact process of geospatial clustering was repeated to get a detailed understanding of the local areas and their built similarity. The map shows a more detailed overview of neighborhood-level building groups and ratio estimation. Since the tiled structure does not follow the administratively defined borders, it aims for a more neutral and practical output. However, it is to note that aggregation of values over grids or any such pattern still amounts to a Modifiable Aerial Unit Problem (MAUP), which might provide different results based on the dimensions of the grid or any aggregation method used. Figure 10 shows a few districts form an outlier group with more ratio near the borders of the Berlin state. Upon analyzing those locations with grid-based aggregation (Figure 11), it can be seen that they (Figure 11 (1)) consist of areas with monotonous built types, such as allotment gardens, with higher *k* than the neighboring tiles. Contrary to that, the city core of *Charlottenburg-Wilmersdorf* hints at wide diversity in building shapes and typology, as visible in Figure 11 (2). Similarly, Figure 11 (3) shows the example of residential blocks constructed post-war, showing L-shaped housing apartments of varying sizes, which offer slightly higher *k* value, while a mix of warehouses and allotment gardens in Figure 11 (4) shows higher *k* than the previous two examples.

The detailed overview of the built similarity and morphology can be explored widely in a similar manner and might be worthy of a fresh perspective in new research. This research primarily sought to group buildings by adjusting settlement boundaries to achieve nuanced differentiations within the urban landscape. Examining the clustering of buildings is inherently intricate, given the multifaceted factors, including building density, accessibility, population, amenities, and more. Nonetheless, the objective of this research to experiment with the computational methods of grouping buildings showed exciting opportunities. This newfound understanding can be a foundation for diverse practical applications within this domain.

## 5. Discussion

The evolution of complex 3D city models is at the forefront of advances in urban design and research. For example, identifying built characteristics and features helps study the overall spatial distribution and allows us to explore urban areas in detail. While the shape metrics are excellent methods to define built structures, the unsupervised methods excel in capturing latent semantic representations of 3D shapes, encompassing structural, orientational, and abstract features beyond mere geometric properties and enabling enhanced generalization. FoldingNet 3D point cloud-based encoder helped obtain a vector representation of the 3D building structures and process them to identify building clusters. This allowed us to compare the buildings in their entirety without primarily focusing on a few geometrical properties. The autoencoder performs well in approximating the shape, which helped create large-scale comparisons with different districts and local areas in the selected study area.

While providing an effective tool for developing quick estimates of building energy requirements, disaster risk reduction, and building-stock assessment, the study also paves the way for future exciting applications in understanding urbanization processes and larger-scale building patterns, particularly regarding the history and newer developments. Instead of studying only the growth pattern in urban regions, it can be worthwhile to map specific forms and functions of built forms that developed over a particular period. For instance, specific historical events in the selected study region, such as the German reunification and the following spatial development. When assisted by other relevant information, such as demographics and building materials, this attempt to group buildings might help locate regions with deprivation and poverty. Apart from the inherent characteristics of buildings, the building age might give insights into changes in design and planning methods and features of the era in which it was built. Although this study only focuses on buildings, other 3D urban and green infrastructure, such as bridges, streets, and trees, can be analyzed with the same method with minimal model modifications. Further, studying the nearness of building units to other urban infrastructures and building groups to each other would be intriguing. Studying the landscape in each building's proximity potentially enables us to answer questions like has the design of the building been affected by the surrounding landscape and/or the location? Or what does the average building neighborhood in the region look like? Lidar datasets and VHR aerial imagery datasets with DL classification models would potentially help locate and map buildings, trees, open spaces, and streets. Similar to this study, the proximity can be encoded into embeddings to comment on the design of urban and rural neighborhoods in the study area.

Although the methodology to group buildings based on the autoencoder model offers flexibility in research and analysis not limited to only urban-based analysis, a few limitations restrict its full capability. Firstly, not all CityGML shapes could be converted to point-based representation due to the lack of a watertight 3D model; this removed approximately 30 percent of the buildings, which might provide slightly different results. Although we are not employing these models for shape metrics calculations but solely for generating point samples, we could still utilize these buildings in our process; however, we consciously opted against their utilization due to concerns regarding potential shape inaccuracies. Nevertheless, since this percentage is not area-specific but evenly spread throughout the study area, the final analysis's effect might be minimal. Secondly, the built structures can be highly nuanced, mainly if we use this methodology in CityGML shapes greater than LoD 2, which have openings for windows and doors with interior room segregation. Currently, the ability of the FoldingNet autoencoder is not checked against highly complex datasets. Also, the conversion of shapes into the point cloud loses precise information regarding edges, which are the prime factors in determining overall shapes, especially in the case of buildings. Increasing the sample point size from currently 2048 points might further improve the learning and reconstruction of complex shapes with a downside of higher computation power requirement. Finally, as with all the autoencoders, the embeddings generated suffer from a lack of contextual information on the datasets on which they are generated. They have limited generalization to unseen dataset samples; for example, generating embeddings for buildings from different regions or countries will differ significantly. The dependency on pre-training data is vast, and the reconstruction quality will depend on the pre-trained dataset's size and representativeness.

## 6. Conclusion

This study delved into the innovative use of utilizing latent vectors derived from LoD 2 CityGML 3D shapes to facilitate shape grouping via similarity metrics. This approach presents a departure from conventional methods and offers promising results. Our findings highlight the potential of latent dimensions as effective approximations for representing 3D shapes, surpassing the need for traditional geometric metrics. Leveraging latent vectors generated by FoldingNet, applied to an extensive dataset encompassing 2 million structures in the German region of Brandenburg and Berlin, our study introduces a geospatial model capable of clustering similar structures through cosine similarity metrics.

This clustering methodology allows in-depth analysis of urbanization patterns, core city development, and historical evolution. Although this study has centered on LoD 2 CityGML shapes characterized by generalized roof shapes and attached walls, it is worth noting that the approach remains adaptable to architecturally detailed LoD 3 and other urban 3D datasets. This adaptability allows for a more nuanced comparison of structures without necessitating substantial changes to the processing pipeline. This research provides a novel perspective on 3D shape representation and lays the foundation for comprehensive urban development studies that can drive informed decision-making in urban planning and development.